\documentclass[prb,superscriptaddress,showpacs,floatfix,onecolumn]{revtex4}
\usepackage{amssymb}
\usepackage{amsmath}
\usepackage{graphicx}

\begin{document}

\title{Dynamics of electrons in the quantum Hall bubble phases }
\author{R. C\^{o}t\'{e}}
\affiliation{D\'{e}partement de physique, Universit\'{e} de Sherbrooke, Sherbrooke, Qu%
\'{e}bec, Canada, J1K 2R1}
\author{C. Doiron}
\affiliation{D\'{e}partement de physique, Universit\'{e} de Sherbrooke, Sherbrooke, Qu%
\'{e}bec, Canada, J1K 2R1}
\author{J. Bourassa}
\affiliation{D\'{e}partement de physique, Universit\'{e} de Sherbrooke, Sherbrooke, Qu%
\'{e}bec, Canada, J1K 2R1}
\author{H. A. Fertig}
\affiliation{Department of Physics and Astronomy, University of Kentucky, Lexington KY
40506-0055}

\begin{abstract}
In Landau levels $N>1$, the ground state of the two-dimensional electron gas
(2DEG)\ in a perpendicular magnetic field evolves from a Wigner crystal for
small filling $\nu ^{\ast }$ of the partially filled Landau level, into a
succession of bubble states with increasing number of guiding centers per
bubble as $\nu ^{\ast }$ increases, to a modulated stripe state near $\nu
^{\ast }=0.5$. In this work, we show that these first-order phase
transitions between the bubble states lead to measurable discontinuities in
several physical quantities such as the density of states and the
magnetization of the 2DEG. We discuss in detail the behavior of the
collective excitations of the bubble states and show that their spectra have
higher-energy modes besides the pinned phonon mode. The frequencies of these
modes, at small wavevector $\mathbf{k}$, have a discontinuous evolution as a
function of filling factor that should be measurable in, for example,
microwave absorption experiments.
\end{abstract}

\date{\today}
\maketitle

\section{Introduction}

It is now well established both theoretically \cite{stripesdebut} and
experimentally\cite{lilly},\cite{du} that the two-dimensional electron gas
(2DEG) ground state near half filling in the higher Landau levels ($N>1$) is
the quantum Hall stripe state. Transport experiments have shown that this
stripe state has an highly anisotropic longitudinal conductivity i.e. $%
\sigma _{yy}>>\sigma _{xx}$ (where $\widehat{x}$ is the direction
perpendicular to the stripes) and a Hall conductivity $\sigma _{xy}$ that is
not quantized. Away from half filling, more precisely at \textit{partial}
filling factors around $\nu ^{\ast }\approx 1/4$ and $\nu ^{\ast }\approx
3/4 $ in Landau level $N=2$, the isotropy of the longitudinal conductivity
is restored and a minima appears in the diagonal resistance. Concomitantly,
the Hall conductivity is quantized, but at a value equal to that of the
nearest integer quantum Hall effect plateaus. These other ground states
around $\nu ^{\ast }\approx 1/4$ and $\nu ^{\ast }\approx 3/4$ in $N=2$ have
been called reentrant integer quantum hall states (RIQHS) \cite{cooper}.
Studies by Koulakov, Fogler, and Shklovskii and Moessner and Chalker \cite%
{stripesdebut} suggest that the RIQHS are due to the formation of bubble
states. These states can be described as triangular Wigner crystals of
electron clusters. Each cluster (or ``bubble'') contains a fixed number $M$
of electrons such that, within them, the local filling factor is equal to
one. The bubble states, being pinned by disorder, should be insulating in
contrast to the stripe states. A review of these novel ground states in
higher Landau level is given in Ref. \onlinecite{foglerreview}.

The ground state of the 2DEG changes discontinuously as the partial filling
factor $\nu ^{\ast }$ is changed. In the Hartree-Fock approximation (HFA),
the ground state is a Wigner crystal at small filling $\nu ^{\ast }$ or,
equivalently, a bubble state with one guiding center per bubble. As the
partial filling is increased, bubble states with increasing number $M$ of
electrons per bubble are stabilized so that the 2DEG evolves from the Wigner
crystal, through a succession of bubble states, into finally the stripe
state near half filling. In Landau level $N$, it can be shown that (in the
HFA) the last bubble state has $M=N+1$. The optimal number of guiding
centers in a bubble is well approximated by the formula $M=3\nu ^{\ast }N$
corresponding to an average separation between the bubbles equal to $%
a=3.3R_{c}$ where $R_{c}=\sqrt{2N+1}\ell $ is the cyclotron radius and $\ell
=\sqrt{\hslash c/eB}$ is the magnetic length. This sequence of phase
transitions in higher Landau levels has been verified by a number of
authors. By comparing the energy of the Laughlin liquid at $\nu ^{\ast }=1/3$
and $\nu ^{\ast }=1/5$ with that of the bubble states in several Landau
levels, Fogler and Koulakov\cite{fogler1} showed, within the HFA, that the
bubble states are lower in energy than the Laughlin liquid at filling factor 
$\nu ^{\ast }=1/3$ for $N\geq 2$ and at filling factor $\nu ^{\ast }=1/5$
for $N\geq 3$. Rezayi et al. \cite{rezayi} used exact diagonalization
studies on finite size systems to show the absence of fractional quantum
hall states (FHQS) in higher Landau levels and a tendancy to charge density
wave (CDW) ordering. Recent calculations by Yoshioka and Shibata\cite%
{yoshioka} using the density matrix renormalization group method (DMRG)
basically confirms the Hartree-Fock scenario with some minor corrections.
For $N=2$ and $N=3$, the DMRG gives a wider region of stability for the
stripe phase than in the HFA. As a result, the last bubble state (that with $%
M=N+1$) is absent from the phase diagram. The transitions between different
bubble states occur at slightly different filling factors in the two
approaches. Both calculations, however, give first order phase transitions
between the bubble states and between the bubble and stripe states. The
possibility of a continuous phase transition involving the deformation of
the bubbles from a isotropic shape to a more elliptical shape as the stripe
state is approached was investigated (with a negative result) by Ren et al. 
\cite{ren} within the HFA. These authors also claimed\cite{yang} that the $%
\nu ^{\ast }=1/6$ composite fermion state has lower energy than the
corresponding bubble state in $N=2$ and that this liquid state could be an
intermediate state between the Wigner crystal and the $M=2$ bubble state.
(See, however, Ref. \onlinecite{yoshioka}.) In any case, the difference in
the cohesive energy of the various possible ground states at any given
filling factor is usually quite small and the possibility that quantum
fluctuations, disorder, finite extension of the wave function in the third
dimension, screening corrections, \textit{etc}. modify the above scenario
cannot be ruled out.

The existence of the bubble state near filling factors $\nu ^{\ast }\approx
1/4$ and $\nu ^{\ast }\approx 3/4$ in $N=2$, was first evidenced by
transport measurements showing its insulating behavior\cite{lilly},\cite{du}
and its nonlinear $I-V$ properties\cite{cooper}. In more recent experiments,
Lewis and al.\cite{lewis} and Ye et al.\cite{ye} measured the diagonal
conductivity $\sigma _{xx}$ of the 2DEG in the microwave regime and found
sharp resonances in $\Re\left[ \sigma _{xx}\right] $ at some frequency.
These resonances are strongest near $\nu ^{\ast }\approx 1/4$ and $\nu
^{\ast }\approx 3/4$ in level $N=2.$ They appear for a range of partial
filling factors around $\nu ^{\ast }\approx 1/4$ and $\nu ^{\ast }\approx
3/4 $ and only for temperatures below $0.1$ K. It is believed that these
resonances are due to the excitation of the pinning mode of the bubble
states realized at these filling factors. The experiments show no hint of
discontinuous transitions among the different bubble states. It is possible
that thermal or disorder effects may smear out any sharp effects on the
pinning mode.

In light of these microwave experiments, it is natural to ask how one could
observe a transition between bubble states, and also how one could
distinguish between bubble states of different $M$ numbers. In this paper,
we study how the behavior of several measurable physical properties of the
bubble states evolve as the filling factor, $\nu ^{\ast }$, is increased
from the Wigner crystal state at low filling to the stripe state at half
filling. We show that the series of transitions between bubble states lead
to oscillations and discontinuity in the orbital magnetization and
susceptibility and to features in the one-particle density of states that
are directly related to the number of electrons per bubble. Working in the
time-dependent Hartree-Fock approximation (TDHFA), we also compute the
dispersion relation of the phonon as well as those of other higher-energy
modes that correspond to excitations localized on the bubbles. These
localized excitations are interesting because (in contrast to the phonon
mode) they are gapped and probably less sensitive to the disordering
potential. Their dispersion can be computed, in a first approximation, by
neglecting disorder which is difficult to take into account in these charge
density wave states. We thus apply the TDHFA to show that, within the range
of frequencies accessible in microwave experiments, the first order
transitions between the bubble states can be tracked by looking at the
discontinuous behavior of the first few high-energy modes as a function of
partial filling factor.

The outline of the paper is as follows. In Sec. II, we briefly describe the
Hartree-Fock procedure to compute the order parameters of the charge density
wave states as well as the orbital magnetization and the one-particle
density of states. We then introduce the time-dependent Hartree-Fock
approximation and describe how we get the dispersion relation of the
collective excitations in these ordered states. In Sec. III, we discuss our
numerical results for these quantities. We conclude in Sec. IV, by
discussing how our results could be related to the experiments of Lewis et
al. and Ye et al..

\section{Hartree-Fock formalism}

\subsection{Order parameters}

The bubble phase is a crystal state that can be described by the set of
average values $\left\{ \left\langle n\left( \mathbf{G}\right) \right\rangle
\right\} $ where $\left\langle n\left( \mathbf{G}\right) \right\rangle $ is
the ground-state average of the density operator and $\mathbf{G}$ is a
reciprocal lattice vector of the crystal. It is convenient, in the Landau
gauge, to define a related density operator by the expression 
\begin{equation}
\rho (\mathbf{G})=\frac{1}{N_{\varphi }}\sum_{X}e^{-iG_{x}X+iG_{x}G_{y}l_{%
\bot }^{2}/2}\ c_{N,X}^{\dagger }c_{N,X-G_{y}l_{\bot }^{2}},
\end{equation}%
where $c_{N,X}^{\dagger }$ creates an electron in Landau level $N$ with
guiding center quantum number $X$ and $N_{\varphi }$ is the Landau level
degeneracy. This new density operator is related to the real density by the
expression 
\begin{equation}
n(\mathbf{G})=N_{\varphi }F_{N,N}(\mathbf{G})\rho (\mathbf{G}),  \label{a_1}
\end{equation}%
where $F_{N,N}(\mathbf{G})$ is a form factor for the $N$ th Landau level
which is given by 
\begin{equation}
F_{N,N}(\mathbf{G})=e^{-G^{2}\ell ^{2}/4}L_{N}^{0}\left( \frac{G^{2}\ell ^{2}%
}{2}\right) ,  \label{1_2}
\end{equation}%
and $L_{N}^{0}\left( x\right) $ is a generalized Laguerre polynomial. At a
semiclassical level, the average $\left\langle \rho (\mathbf{G}%
)\right\rangle $ can be viewed as a Fourier transform of a ``guiding center
density'' of cyclotron orbits.

In this paper, we make the usual approximations of neglecting any Landau
level mixing and consider that the filled Landau levels are inert. It
follows that the filled levels do not enter in our calculation and we will,
whenever possible, drop the level index $N$ from now on to simplify the
notation. Moreover, we assume that the partially filled level is completely
spin polarized.

The average values $\left\langle \rho (\mathbf{G})\right\rangle $ are
obtained by computing the single-particle Green's function 
\begin{equation}
G\left( X,X^{\prime },\tau \right) =-\left\langle Tc_{X}\left( \tau \right)
c_{X^{\prime }}^{\dagger }\left( 0\right) \right\rangle ,
\end{equation}%
whose Fourier transform we define as 
\begin{equation}
G\left( \mathbf{G,}\tau \right) =\frac{1}{N_{\phi }}\sum_{X,X^{\prime }}e^{-%
\frac{i}{2}G_{x}\left( X+X^{\prime }\right) }\delta _{X,X^{\prime
}-G_{y}l_{\bot }^{2}}G\left( X,X^{\prime },\tau \right) ,  \label{1_3}
\end{equation}%
so that 
\begin{equation}
\left\langle \rho \left( \mathbf{G}\right) \right\rangle =G\left( \mathbf{G,}%
\tau =0^{-}\right) .
\end{equation}

We compute the single-particle Green's function of Eq. (\ref{1_3}) by
solving numerically the Hartree-Fock equation of motion%
\begin{equation}
\left[ i\omega _{n}+\mu \right] G\left( \mathbf{G,}i\omega _{n}\right)
-\sum_{\mathbf{G}^{\prime }}U\left( \mathbf{G}-\mathbf{G}^{\prime }\right)
e^{i\mathbf{G}\times \mathbf{G}^{\prime }\ell ^{2}/2}G\left( \mathbf{G}%
^{\prime }\mathbf{,}i\omega _{n}\right) =\delta _{\mathbf{G},0},  \label{1_4}
\end{equation}%
where $\omega _{n}$ is a fermionic Matsubara frequency and the Hartree-Fock
potential $U\left( \mathbf{q}\right) $ is given by%
\begin{equation}
U\left( \mathbf{q}\right) =\left[ H_{N,N}\left( \mathbf{q}\right)
-X_{N,N}\left( \mathbf{q}\right) \right] \left\langle \rho \left( -\mathbf{q}%
\right) \right\rangle .  \label{1_17}
\end{equation}%
In Eq. (\ref{1_17}), the Hartree and Fock interactions are given by 
\begin{eqnarray}
H_{N,N}\left( \mathbf{q}\right) &=&\left( \frac{e^{2}}{\kappa \ell }\right) 
\frac{1}{q\ell }e^{\frac{-q^{2}\ell ^{2}}{2}}\left[ L_{N}^{0}\left( \frac{%
q^{2}\ell ^{2}}{2}\right) \right] ^{2}, \\
X_{N,N}\left( \mathbf{q}\right) &=&\left( \frac{e^{2}}{\kappa \ell }\right) 
\sqrt{2}\int_{0}^{\infty }dx\,e^{-x^{2}}\left[ L_{N}^{0}\left( x^{2}\right) %
\right] ^{2}J_{0}\left( \sqrt{2}xq\ell \right) ,
\end{eqnarray}%
where $\kappa $ is the dielectric constant of the host semiconductor and $%
J_{0}\left( x\right) $ is the Bessel function of order zero.

The electronic density in real space is obtained by the relation

\begin{equation}
\left\langle n\left( \mathbf{r}\right) \right\rangle =\frac{1}{2\pi \ell ^{2}%
}\sum_{\mathbf{G}}e^{+i\mathbf{G}\cdot \mathbf{r}}F_{N,N}\left( \mathbf{G}%
\right) \left\langle \rho \left( \mathbf{G}\right) \right\rangle .
\label{1_9}
\end{equation}

\subsection{Hartree-Fock energy}

We solve the Hartree-Fock equation of motion (Eq. (\ref{1_4})) numerically
by using an iterative approach that was described in detail in Refs. %
\onlinecite{cote1} . Once the order parameters$\left\{ \left\langle \rho
\left( \mathbf{G}\right) \right\rangle \right\} $ are found, the
Hartree-Fock energy per particle in the partially filled level can be
written as 
\begin{equation}
\frac{E_{HF}}{N_{s}}={\frac{1}{{2}\nu ^{\ast }}}\sum_{\mathbf{G}}\left[
H_{N,N}\left( \mathbf{G}\right) \left( 1-\delta _{\mathbf{G},0}\right)
-X_{N,N}\left( \mathbf{G}\right) \right] \left| \left\langle \rho \left( 
\mathbf{G}\right) \right\rangle \right| ^{2},  \label{1_14}
\end{equation}%
where $\nu ^{\ast }$ is the filling factor of Landau level $N$ and $N_{s}$
is the number of electrons in the partially filled level. The total filling
factor of the two-dimensional electron gas is $\nu =2N+\nu ^{\ast }$.

\subsection{Density of states}

In our iterative approach, the single-particle Green's function is computed
by first finding the eigenvalues and eigenvectors of the interaction matrix 
\begin{equation}
F_{\mathbf{G},\mathbf{G}^{\prime }}\equiv U\left( \mathbf{G}-\mathbf{G}%
^{\prime }\right) e^{i\mathbf{G}\times \mathbf{G}^{\prime }\ell ^{2}/2}.
\end{equation}%
This matrix is hermitian and can be diagonalized with the unitary
transformation%
\begin{equation*}
F=VDV^{\dagger },
\end{equation*}%
where $V$ is the matrix of the eigenvectors of $F$ and $D_{i,j}=d_{j}\delta
_{i,j}$ is the diagonal matrix of the eigenvalues of $F$. The Green's
function is then readily given by%
\begin{equation}
G\left( \mathbf{G,}\omega _{n}\right) =\sum_{j}\frac{V_{\mathbf{G},j}\left[
V^{\dagger }\right] _{j,\mathbf{G}=0}}{i\omega _{n}+\left( \mu -d_{j}\right)
/\hslash }.
\end{equation}

The density of states is defined as%
\begin{equation}
g\left( \omega \right) =-\frac{1}{\pi }\int d\mathbf{r}\,\Im\left[
G^{R}\left( \mathbf{r},\mathbf{r},\omega \right) \right] ,
\end{equation}%
where $G^{R}$ is the retarded single-particle Green's function which can be
computed from the eigenvalues and eigenvectors found above. We then have

\begin{equation}
g\left( \omega \right) =-\frac{N_{\phi }}{\pi }\Im\left[ \sum_{j}\frac{%
\left| V_{\mathbf{G}=0,j}\right| ^{2}}{\omega +i\delta -d_{j}/\hslash }%
\right] .  \label{1_11}
\end{equation}

\subsection{Collective modes}

As shown in Refs. \onlinecite{cote1}, the order parameters $\left\{
\left\langle \rho \left( \mathbf{G}\right) \right\rangle \right\} $ can also
be used to compute the density-density response function $\chi _{\mathbf{G},%
\mathbf{G}^{\prime }}\left( \mathbf{k},\tau \right) $ defined by%
\begin{equation}
\chi _{\mathbf{G},\mathbf{G}^{\prime }}\left( \mathbf{k},\tau \right)
=-N_{\varphi }\left\langle T\widetilde{\rho }\left( \mathbf{k}+\mathbf{G}%
,\tau \right) \widetilde{\rho }\left( -\mathbf{k}-\mathbf{G}^{\prime
},0\right) \right\rangle ,
\end{equation}%
where $\widetilde{\rho }\equiv \rho -\left\langle \rho \right\rangle $. The
collective modes of the bubble state are found from the poles of this
response function. By following the poles with non-vanishing weight as the
wavevector $\mathbf{k}$ is varied in the Brillouin zone of the reciprocal
lattice, we get the dispersion relation of the phonon and higher-energy
collective modes.

The equation of motion for the density response in the TDHFA is given by the
matrix equation 
\begin{equation}
\sum_{\mathbf{G}^{\prime \prime }}\left[ i\Omega _{n}\delta _{\mathbf{G},%
\mathbf{G}^{\prime }}-M_{\mathbf{G},\mathbf{G}^{\prime \prime }}\left( 
\mathbf{k}\right) \right] \chi _{\mathbf{G}^{\prime \prime },\mathbf{G}%
^{\prime }}\left( \mathbf{k},i\Omega _{n}\right) =B_{\mathbf{G},\mathbf{G}%
^{\prime }}\left( \mathbf{k}\right) ,  \label{7p1}
\end{equation}%
where $\Omega _{n}$ is a Matsubara bosonic frequency and the matrices $M_{%
\mathbf{G},\mathbf{G}}$ and $B_{\mathbf{G},\mathbf{G}^{\prime }}$ are
defined by 
\begin{eqnarray}
M_{\mathbf{G},\mathbf{G}^{\prime }}\left( \mathbf{k}\right) &=&-2i\left( 
\frac{e^{2}}{\hslash \kappa \ell }\right) \left\langle \rho \left( \mathbf{%
G-G}^{\prime }\right) \right\rangle \\
&&\times \sin \left[ \frac{\left( \mathbf{G}\times \mathbf{G}^{\prime
}\right) \ell ^{2}}{2}\right] \left[ H_{N,N}\left( \mathbf{G}-\mathbf{G}%
^{\prime }\right) -X_{N,N}\left( \mathbf{G-G}^{\prime }\right)
-H_{N,N}\left( \mathbf{G}^{\prime }\right) +X_{N,N}\left( \mathbf{G}^{\prime
}\right) \right] ,  \notag
\end{eqnarray}%
\newline
and%
\begin{equation}
B_{\mathbf{G},\mathbf{G}^{\prime }}\left( \mathbf{k}\right) =2i\sin \left[ 
\frac{\left( \mathbf{G}\times \mathbf{G}^{\prime }\right) \ell ^{2}}{2}%
\right] \left\langle \rho \left( \mathbf{G-G}^{\prime }\right) \right\rangle
\end{equation}%
respectively.

To solve for $\chi _{\mathbf{G},\mathbf{G}^{\prime }}\left( \mathbf{k}%
,i\Omega _{n}\right) $, we diagonalize the matrix $M_{\mathbf{G},\mathbf{G}%
^{\prime \prime }}\left( \mathbf{k}\right) $ by the transformation

\begin{equation}
M=WTW^{-1},
\end{equation}%
where $W$ is the matrix of the eigenvectors of $M$ and $T_{i,j}=t_{j}\delta
_{i,j}$ is the diagonal matrix of its eigenvalues. The analytic continuation
of $\chi _{\mathbf{G},\mathbf{G}^{\prime }}\left( \mathbf{k},i\Omega
_{n}\right) $ is given by 
\begin{equation}
\chi _{\mathbf{G},\mathbf{G}^{\prime }}\left( \mathbf{k},\omega \right)
=\sum_{j,k}\frac{W_{\mathbf{G},j}\left( \mathbf{k}\right) \left[ W\left( 
\mathbf{k}\right) ^{-1}\right] _{j,k}B_{k,\mathbf{G}^{\prime }}\left( 
\mathbf{k}\right) }{\omega +i\delta -t_{j}\left( \mathbf{k}\right) }.
\label{1_12}
\end{equation}%
The $j$th eigenvector in the matrix $W$ of the eigenvectors of the matrix $M$
gives the Fourier transform of the density modulation associated with the $j$%
th eigenvalue of $M$. We can thus produce an animation of the motion of the
density modulation with the frequency $t_{j}$ and wave vector $\mathbf{k}$
by computing the time-dependent density%
\begin{equation}
\delta n\left( \mathbf{r},t\right) \sim e^{-it_{j}t}\sum_{\mathbf{G}%
}e^{i\left( \mathbf{k}+\mathbf{G}\right) \cdot \mathbf{r}}F_{N,N}\left( 
\mathbf{k+G}\right) W_{\mathbf{G},j}\left( \mathbf{k}\right)
\end{equation}%
at several values of $t$ in one period of the motion and then superimposing
this density to the ground state density given by the Hartree-Fock
calculation (eq. (\ref{1_9})). We thus compute $n\left( \mathbf{r},t\right)
=n_{HF}\left( \mathbf{r}\right) +\alpha \delta n\left( \mathbf{r},t\right) $
where $\alpha $ is choosen sufficiently small to satisfy $\left| \delta
n\left( \mathbf{r},t\right) \right| <n_{HF}\left( \mathbf{r}\right) $.

\subsection{Orbital magnetisation and magnetic susceptibility}

The orbital magnetisation and susceptibility can be obtained from the
dependence of the Hartree-Fock energy with filling factor. At $T=0K$, the
orbital magnetisation per electron is given by%
\begin{equation}
m=-\frac{1}{N}\left( \frac{\partial E}{\partial B}\right) ,
\end{equation}%
while the orbital magnetic susceptibility per particle is%
\begin{equation}
\chi =\left( \frac{\partial m}{\partial B}\right) ,
\end{equation}%
where $E$ is the ground state energy of the 2DEG.

When all Landau levels are taken into account, the Hartree-Fock energy per
particle, excluding the Zeeman contribution, is (for $0\leq \nu ^{\ast }\leq
1$)

\begin{eqnarray}
\frac{E_{HF}^{TOTAL}}{N_{T}} &=&\frac{1}{\nu }\sum_{M<N}\sum_{\alpha }\left(
M+\frac{1}{2}\right) \hslash \omega _{c}+\frac{\nu ^{\ast }}{\nu }\left( N+%
\frac{1}{2}\right) \hslash \omega _{c}  \label{1_5} \\
&&-\frac{1}{2\nu }\left( \frac{e^{2}}{\kappa \ell }\right) \sum_{\alpha
}\sum_{M<N}\sum_{M^{\prime }<N}X_{M,M^{\prime }}\left( \mathbf{G}=0\right) 
\notag \\
&&-\frac{1}{\nu }\left( \frac{e^{2}}{\kappa \ell }\right)
\sum_{M<N}X_{M,N}\left( \mathbf{G}=0\right) \nu ^{\ast }  \notag \\
&&+\frac{1}{2\nu }\left( \frac{e^{2}}{\kappa \ell }\right) \sum_{\mathbf{G}}%
\left[ H_{N,N}\left( \mathbf{G}\right) \left( 1-\delta _{\mathbf{G}%
,0}\right) -X_{N,N}\left( \mathbf{G}\right) \right] \left| \left\langle \rho
\left( \mathbf{G}\right) \right\rangle \right| ^{2},  \notag
\end{eqnarray}%
where 
\begin{equation}
X_{N,M}\left( \mathbf{q}\right) =\left( \frac{\min \left( M,N\right) !}{\max
\left( M,N\right) !}\right) \int_{0}^{+\infty }dy\left( \frac{y^{2}}{2}%
\right) ^{\left| N-M\right| }e^{-y^{2}/2}\left[ L_{\min \left( N,M\right)
}^{\left| N-M\right| }\left( \frac{y^{2}}{2}\right) \right] ^{2}J_{0}\left(
q\ell _{\perp }y\right) ,
\end{equation}%
is the exchange interaction between electrons in Landau levels $N$ and $M,$
and $\alpha $ is the spin index. To derive this formula, we have assumed
that all levels below $N$ are occupied and that only the spin level $\alpha
=-1$ in Landau level $N$ is partially occupied. The \textit{total} number of
particles $N_{T}$ is assumed fixed.

With the definitions%
\begin{eqnarray}
\Lambda _{0}\left( N\right) &=&\sum_{M<N}\left( M+\frac{1}{2}\right) , \\
\Lambda _{1}\left( N\right) &=&\sum_{M<N}\sum_{M^{\prime }<N}X_{M,M^{\prime
}}\left( \mathbf{G}=0\right) , \\
\Lambda _{2}\left( N\right) &=&\sum_{M<N}X_{M,N}\left( \mathbf{G}=0\right) ,
\\
F\left( N,\nu ^{\ast }\right) &=&\frac{1}{2\nu ^{\ast }}\sum_{\mathbf{G}}%
\left[ H_{N,N}\left( \mathbf{G}\right) \left( 1-\delta _{\mathbf{G}%
,0}\right) -X_{N,N}\left( \mathbf{G}\right) \right] \left| \left\langle \rho
\left( \mathbf{G}\right) \right\rangle \right| ^{2},
\end{eqnarray}%
Eq. (\ref{1_5}) can be written as%
\begin{eqnarray}
\frac{E_{HF}^{TOTAL}}{N_{T}} &=&2\Lambda _{0}\left( N\right) \frac{\hslash
\omega _{c}}{\nu }+\left( N+\frac{1}{2}\right) \frac{\nu ^{\ast }}{\nu }%
\hslash \omega _{c}  \label{1_13} \\
&&-\frac{1}{\nu }\left( \frac{e^{2}}{\kappa \ell }\right) \Lambda _{1}\left(
N\right) -\frac{\nu ^{\ast }}{\nu }\left( \frac{e^{2}}{\kappa \ell }\right)
\Lambda _{2}\left( N\right) +\frac{\nu ^{\ast }}{\nu }\left( \frac{e^{2}}{%
\kappa \ell }\right) F\left( \nu ^{\ast }\right) ,  \notag
\end{eqnarray}%
where%
\begin{equation}
F\left( \nu ^{\ast }\right) ={\frac{1}{{2}\nu ^{\ast }}}\sum_{\mathbf{G}}%
\left[ H_{N,N}\left( \mathbf{G}\right) \left( 1-\delta _{\mathbf{G}%
,0}\right) -X_{N,N}\left( \mathbf{G}\right) \right] \left| \left\langle \rho
\left( \mathbf{G}\right) \right\rangle \right| ^{2},  \label{1_19}
\end{equation}%
is the (dimensionless) Hartree-Fock energy per particle in the partially
filled level i.e. the energy given in Eq. (\ref{1_14}).

Differentiating Eq. (\ref{1_13}) once (twice) with respect to the magnetic
field at constant density and adding the spin contribution, we get the total
magnetization (total spin susceptibility). We write%
\begin{equation}
\mu =\mu _{B}\left( \frac{m}{m^{\ast }}\right) \left( \mu _{1}+\mu _{2}+\mu
_{3}+\mu _{4}\right) ,
\end{equation}%
where $\mu _{B}=e\hslash /2mc$ is the Bohr magneton. The four contributions
to the magnetization are defined (for $0\leq \nu ^{\ast }\leq 1$) by 
\begin{eqnarray}
\mu _{1} &=&-8\Lambda _{0}\frac{1}{\nu }-2\left( N+\frac{1}{2}\right) \left(
1-\frac{4N}{\nu }\right) ,  \label{1_18} \\
\mu _{2} &=&\frac{1}{2}g^{\ast }\left( \frac{m^{\ast }}{m}\right) \left( 1-%
\frac{4N}{\nu }\right) ,  \notag \\
\mu _{3} &=&\frac{3}{\sqrt{2}}r_{s}\Lambda _{1}\left( N\right) \frac{1}{%
\sqrt{\nu }}-\sqrt{2}r_{s}\Lambda _{2}\left( N\right) \left[ -\frac{1}{2}\nu
+3N\right] \frac{1}{\sqrt{\nu }},  \notag \\
\mu _{4} &=&\sqrt{2}r_{s}\sqrt{\nu }\left( \nu -2N\right) \frac{\partial
F\left( \nu ^{\ast }\right) }{\partial \nu ^{\ast }}+\sqrt{2}r_{s}\left[ -%
\frac{1}{2}\nu +3N\right] \frac{1}{\sqrt{\nu }}F\left( \nu ^{\ast }\right) .
\notag
\end{eqnarray}%
Similarly, we write (for $0\leq \nu ^{\ast }\leq 1$) the total magnetic
susceptibility as the sum of the four contributions

\begin{equation}
\chi =\frac{e^{2}}{4\pi nm^{\ast }c^{2}}\left( \chi _{1}+\chi _{2}+\chi
_{3}+\chi _{4}\right) ,  \label{1_8}
\end{equation}%
where%
\begin{eqnarray}
\chi _{1} &=&-8\Lambda _{0}\left( N\right) +8N\left( N+\frac{1}{2}\right) ,
\label{1_16} \\
\chi _{2} &=&-2Ng^{\ast }\left( \frac{m^{\ast }}{m}\right) ,  \notag \\
\chi _{3} &=&\frac{3r_{s}}{2\sqrt{2}}\Lambda _{1}\left( N\right) \sqrt{\nu }-%
\frac{1}{\sqrt{2}}r_{s}\Lambda _{2}\left( N\right) \left[ \frac{\nu ^{3/2}}{2%
}+3N\sqrt{\nu }\right] ,  \notag \\
\chi _{4} &=&\sqrt{2}r_{s}\left[ \frac{1}{4}\left( \nu +6N\right) \sqrt{\nu }%
F\left( \nu ^{\ast }\right) -\left( \nu +2N\right) \nu ^{3/2}\frac{\partial
F\left( \nu ^{\ast }\right) }{\partial \nu ^{\ast }}-\left( \nu -2N\right)
\nu ^{5/2}\frac{\partial ^{2}F\left( \nu ^{\ast }\right) }{\partial \nu
^{\ast 2}}\right] .  \notag
\end{eqnarray}%
The four contributions to the magnetization and susceptibility come
respectively from the kinetic energy ($\mu _{1},\chi _{1}$), the Zeeman
energy ($\mu _{2},\chi _{2}$), the exchange energy between electrons in the
filled levels or between electrons in the filled levels and those in level $%
N $ ($\mu _{3},\chi _{3}$) and the Hartree-Fock energy of electrons in level 
$N $ ($\mu _{4},\chi _{4}$). In Eq. (\ref{1_16}), we have introduced the gas
parameter $r_{s}=1/\sqrt{\pi na_{B}^{2}}$where $a_{B}=\hslash ^{2}\kappa
/m^{\ast }e^{2}$ is the effective Bohr radius and the effective $g$ factor $%
g^{\ast }$. We remark that our units for the susceptibility can also be
written as $e^{2}/4\pi nm^{\ast }c^{2}=\mu _{B}^{\ast }/B_{\nu =1}$ where $%
\mu _{B}^{\ast }=e\hslash /2m^{\ast }c$ and $B_{\nu =1}$ is the magnetic
field required to get a filling factor of $\nu =1$ for a total density of $n$%
.

\section{Numerical results}

We now discuss our numerical results for the physical quantities introduced
above.

\subsection{Hartree-Fock energy}

Figures 1(a)-(c) shows the Hartree-Fock energy per electron as a function of
filling factor for the bubble and stripe phases in Landau levels $N=2,3$and $%
5$. The filling factors at which the transitions between states occur are
indicated in an inset in each figure. In all cases, the transitions are of
first order and the transition scenario is that described in our
introduction: the ground state evolves from the Wigner crystal at low
filling, through a succession of bubble states with increasing values of
guiding centers per bubble, and finally into the stripe state near half
filling. The number of possible bubble states increases with Landau level
index $N$. Our results for $N=2,3$are in agreement with previously obtained
Hartree-Fock energies for the bubble states\cite{yoshioka}. Energies for $%
N=5 $ have not been computed before. As shown in Ref. \onlinecite{yoshioka},
the Hartree-Fock results are similar to those obtained by a DMRG calculation
except that the region where the stripe phase is stable is wider in the DMRG
calculation and, consequently, the bubble state with $N+1$electron per
bubble disappears. Oscillations in the Laguerre polynomial entering the form
factor in Eq. (\ref{1_2}) make it difficult to compute numerically accurate
phase diagram for higher Landau level indices using our equation of motion
technique.

\begin{figure}[tbp]
\includegraphics[width=8.25cm]{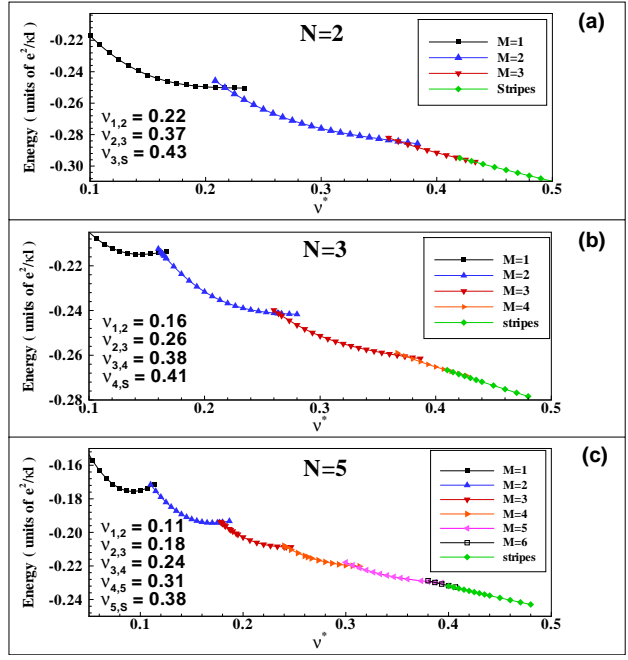}
\caption{Hartree-Fock energy per particle as a function of the filling
factor $\protect\nu ^{\ast }$ of the partially filled level for the bubble
and stripe phases in higher Landau level. The inset in each figure indicates
the critical filling factor $\protect\nu _{i,j}$ of the \textit{partially}
filled level at which the transition between bubble states or between the
last bubble state and the stripe state occurs. }
\label{fig1}
\end{figure}

\subsection{Density pattern}

Figure 2 shows how the density pattern given by Eq. (\ref{1_9}) evolves as
the filling factor increases in Landau level $N=3$. When the bubbles are
relatively far apart, the density pattern on each crystal site is well
approximated by the trial wave function of Fogler and Koulakov\cite{fogler1}%
. For a bubble of $M$ electrons in Landau level $N$, that wave function is
the Slater determinant%
\begin{equation}
\Psi _{N}\left( \mathbf{r}_{1},\mathbf{r}_{2},...,\mathbf{r}_{M}\right)
=\left\vert 
\begin{array}{cccc}
\varphi _{N,0}\left( \mathbf{r}_{1}\right)  & \varphi _{N,0}\left( \mathbf{r}%
_{2}\right)  & ... & \varphi _{N,0}\left( \mathbf{r}_{M}\right)  \\ 
\varphi _{N,1}\left( \mathbf{r}_{1}\right)  & \varphi _{N,1}\left( \mathbf{r}%
_{2}\right)  & ... & \varphi _{N,1}\left( \mathbf{r}_{M}\right)  \\ 
\vdots  & \vdots  & \vdots  & \vdots  \\ 
\varphi _{N,M-1}\left( \mathbf{r}_{1}\right)  & \varphi _{N,M-1}\left( 
\mathbf{r}_{2}\right)  & ... & \varphi _{N,M-1}\left( \mathbf{r}_{M}\right) 
\end{array}%
\right\vert .
\end{equation}%
where $\varphi _{N,m}\left( \mathbf{r}\right) =C_{N,m}\left( \frac{r}{\ell }%
\right) ^{\left\vert m-N\right\vert }e^{-r^{2}/4\ell ^{2}}e^{i\left(
N-m\right) \theta }L_{\left( N+m-\left\vert m-N\right\vert \right)
/2}^{\left\vert m-N\right\vert }\left( \frac{r^{2}}{2\ell ^{2}}\right) $is
the normalised wave function of an electron in the symmetric gauge $\mathbf{A%
}=\left( -B_{0}y/2,B_{0}x/2,0\right) $ with Landau level $N$ and angular
momentum $m$. The one-particle density%
\begin{equation}
n_{N}\left( \mathbf{r}\right) =\left[ \prod_{i=2}^{i=M}\int d\mathbf{r}_{i}%
\right] \left\vert \Psi _{N}\left( \mathbf{r},\mathbf{r}_{2},...,\mathbf{r}%
_{M}\right) \right\vert ^{2},
\end{equation}%
is just 
\begin{equation}
n_{N}\left( \mathbf{r}\right) =\sum_{m=0}^{m=M-1}\left\vert \varphi
_{N,m}\left( \mathbf{r}\right) \right\vert ^{2}.  \label{1_10}
\end{equation}%
As the filling factor increases in the $M$th bubble state, the outer ring
from each bubble gets closer to its neighbors and the density pattern given
by Eq. (\ref{1_10}) is strongly modified. When rings from two adjacent
bubbles start to overlap, there is a transition to the $M+1$th bubble state.
In Landau level $N$, the last bubble state before the transition to the
stripe phase has $M=N+1$. We note that the stripe state obtained in our
Hartree-Fock approximation is not the quantum Hall smectic state. It has
density modulations along the stripes and can be described as an highly
anisotropic Wigner crystal. However, it is likely that this state is
unstable due to quantum fluctuations to the smectic state\cite{fertig}.

\begin{figure}[tbp]
\includegraphics[width=8.25cm]{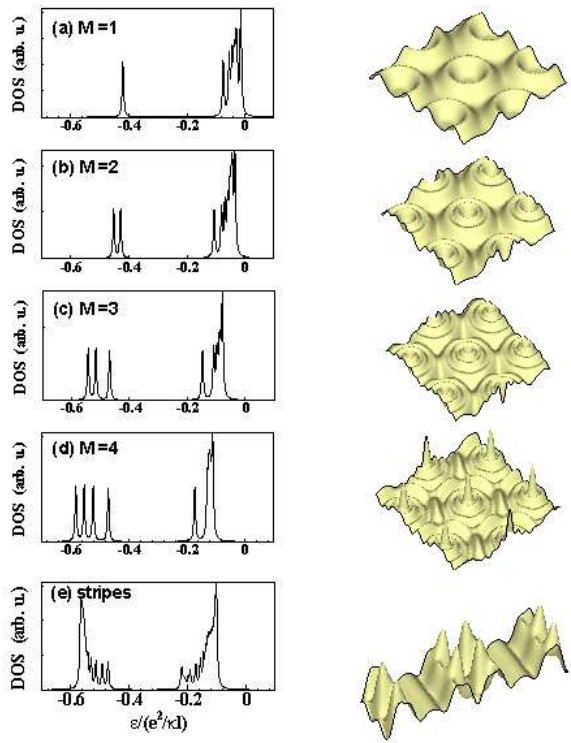}
\caption{Single-particle density of states and density pattern for bubble
state with $M=1,2,3,4$ electrons per bubble and for the stripe states in
Landau level $N=3$. The filling factors are (a) $\protect\nu ^{\ast }=0.113$%
; (b) $\protect\nu ^{\ast }=0.173;$(c) $\protect\nu ^{\ast }=0.320;$ (d) $%
\protect\nu ^{\ast }=0.406;$ (e) $\protect\nu ^{\ast }=0.430$. (See fig2.png
file).}
\label{fig2}
\end{figure}

\subsection{Density of states}

Figure 2 shows the behavior of the single-particle density of states $%
g\left( E\right) $ defined in Eq. (\ref{1_11}) for Landau level $N=3$ and
for filling factors corresponding to bubble states with $M=1,2,3$ and $4$
electrons per bubble. In Landau level $N=0$ where the Wigner crystal state
with $M=1$is the ground state for all filling factor, our calculation for $%
g\left( E\right) $ reproduces the well-known Hofstadter butterfly structure.
For $\nu ^{\ast }=q/p$, the density of states consists of $p$ subbands of
which $q$ low-energy bands are filled and separated by a gap from the
remaining $p-q$ subbands. In Fig. 2, we see that the density of states in
the bubble state has a different structure. The number of low-energy
subbands that can be resolved by our numerical calculation clearly
corresponds to the number $M$ of electrons per bubble. A measurement of the
density of states, by a double-well tunneling experiment or
photoluminescence, could serve to determine $M$. In computing the tunneling
current, one would have to take into account other processes involved,
particularly phonon shake-up. This is however beyond the scope of our paper
and is left for future research. We note for now, however, that the typical
phonon bandwidth (discussed below) is not larger than the splittings among
the occupied bands in the density of states, so that the multiplicities of
peaks associated with different bubble states are likely to survive such
effects.

\subsection{Magnetic susceptibility}

The Hartree-Fock energy shown in Fig. 1 has discontinuous change of slope
and curvature at the transitions between bubble states with different number 
$M$ of electrons per bubble. This, in turn, gives rise to discontinuities in
the behavior of the magnetic susceptibility with filling factor. To show
this, we compute the susceptibility in Landau level $N=3$ by numerically
evaluating the first and second derivatives of the function $F\left( \nu
^{\ast }\right) $defined in Eq. (\ref{1_19}). We use the parameters $\Lambda
_{0}\left( 3\right) =9/2,\Lambda _{1}\left( 3\right) =6.29$ and $\Lambda
_{2}\left( 3\right) =1.32$ appropriate to $N=3$ and assume a typical total
electronic density of $n=3\times 10^{11}$e/cm$^{2}$so that the gas parameter
is $r_{s}=1.011$. The other parameters are $g^{\ast }=0.45$ and the
effective mass $m^{\ast }=0.067m$.

Figures 3(a) and 3(b) show the behavior of the four contributions to the
magnetic moment and magnetic susceptibility defined in Eq. (\ref{1_18}) and
Eq. (\ref{1_16}). The singular behavior (contributions $\mu _{4}$ and $\chi
_{4}$) comes, in both cases, from the Coulomb interaction between electrons
in the partially filled level. The magnetic moment and susceptibility change
discontinuously at the transition between bubble states and between bubble
and stripe state. In both cases the effect is sizeable. Consequently, the
transitions between bubble states should, in theory, be observable in a
measurement of the magnetization or magnetic susceptibility \cite{levy}.

\begin{figure}[tbp]
\includegraphics[width=8.25cm]{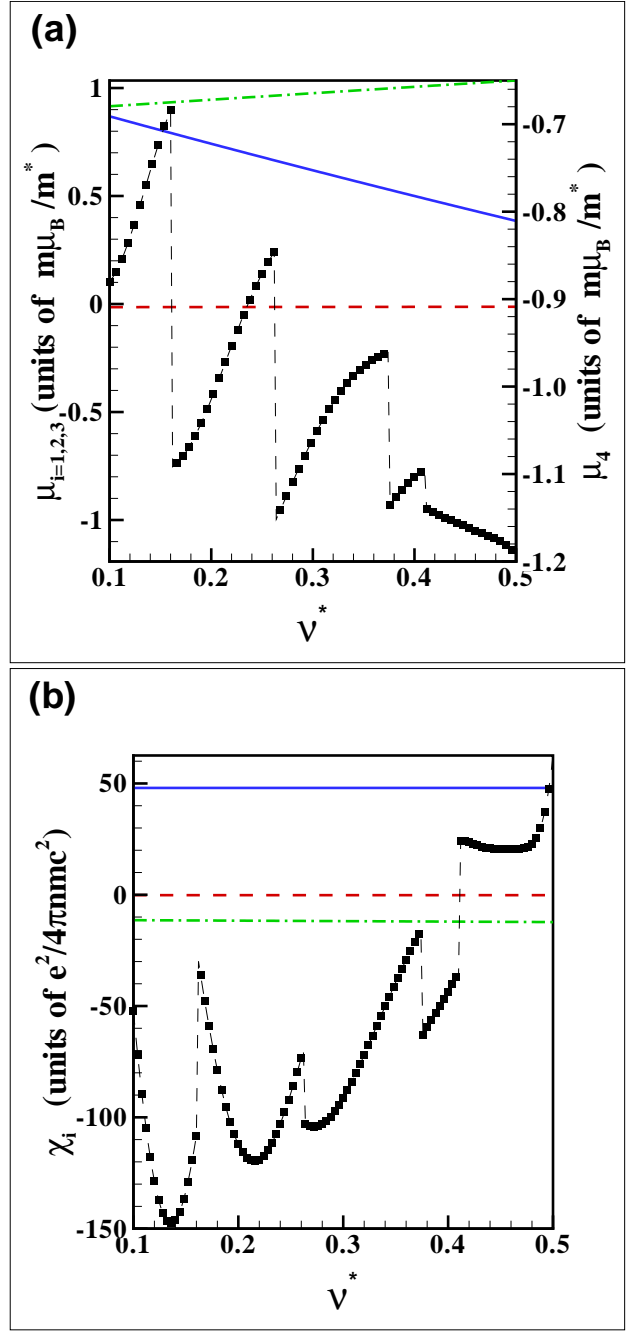}
\caption{(a) Change in the magnetic moment and (b) magnetic susceptibility
with partial filling factor $\protect\nu ^{\ast }$ in Landau level N=3. The
four contributions $i=1$ (solid), $2$ (dashed), $3$ (dashed dot), and $4$
(squares) to $\protect\mu $ or $\protect\chi $ are defined in eqs. (\protect
\ref{1_18}) and (\protect\ref{1_16}). }
\label{fig3}
\end{figure}

\subsection{Collective excitations}

We now discuss the collective excitations of the bubble states. We compute
the density-density response function of Eq. (\ref{1_12}) and follow its
poles when the wave vector $\mathbf{k}$ is varied along the edges of the
irreducible Brillouin zone of the triangular lattice of the crystal. Figures
4(a) to 4(f) show examples of the dispersion relation we obtain. In these
figures, the dispersion is plotted along the path $\Gamma -J-X-\Gamma $
corresponding to the wavevectors $(k_{x},k_{y})=(0,0),\frac{2\pi }{a}(\frac{1%
}{\sqrt{3}},\frac{1}{3}),\frac{2\pi }{a}(\frac{1}{\sqrt{3}},0),\left(
0,0\right) $ where $a$ is the lattice constant. The wavevector $k$
represents the total distance, in reciprocal space and in units of $2\pi /a$
where $a$ is the lattice constant, along the path $\Gamma -J-X-\Gamma $ from
the origin $\Gamma $. For a given value of $\mathbf{k},$ Eq. (\ref{1_13})
provides us with a way to image the motion of the density in a particular
mode.

\begin{figure}[tbp]
\includegraphics[width=8.25cm]{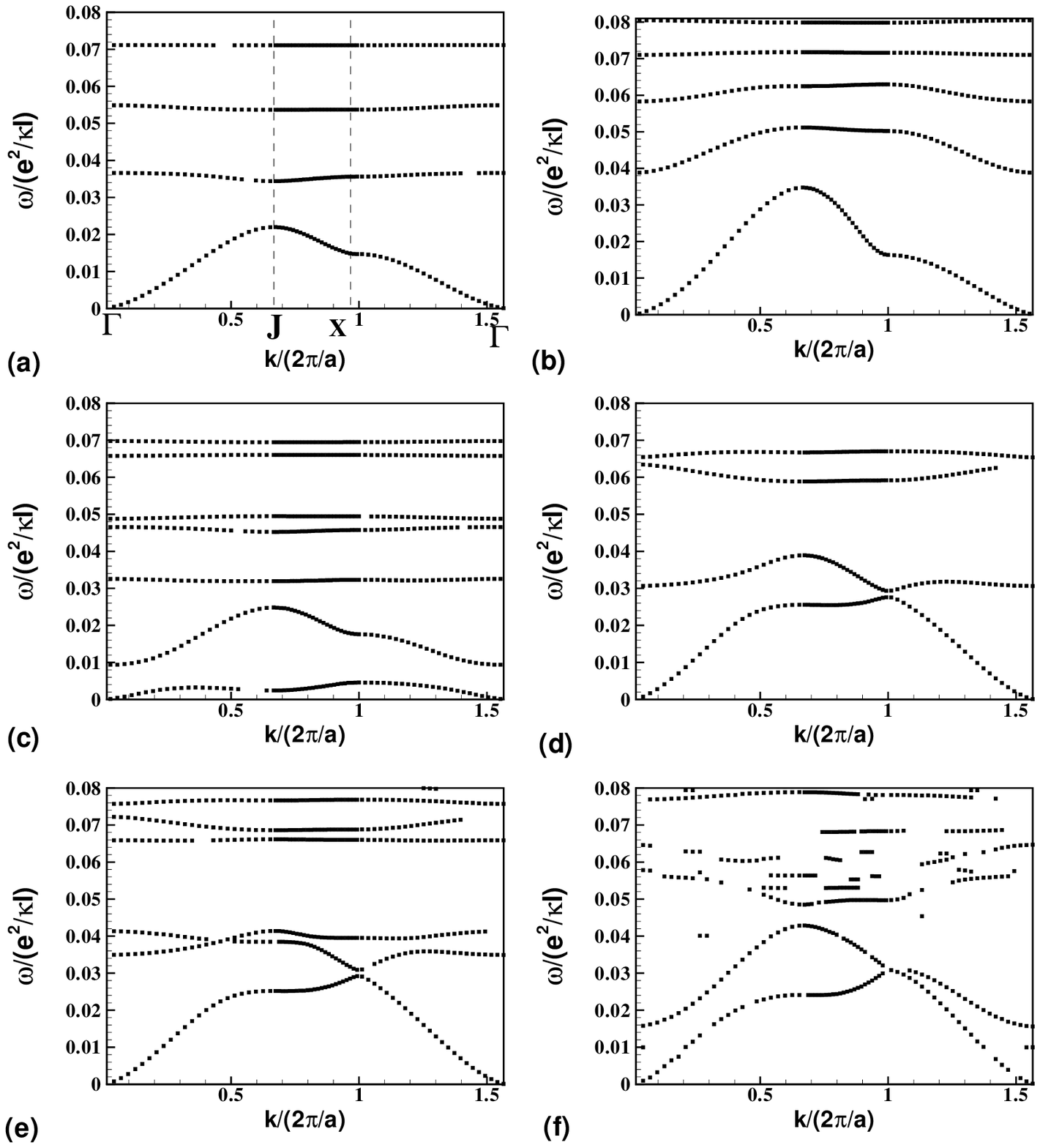}
\caption{Dispersion relations of the phonon and higher-energy modes in
Landau level $N=3$ along the edges (path $\Gamma -J-X-\Gamma $) of the
irreducible Brillouin zone of the triangular lattice.The filling factors and
number of guiding centers per bubble are (a) $\protect\nu ^{\ast }=0.106$, $%
M=1$; (b) $\protect\nu ^{\ast }=0.161$, $M=1$; (c) $\protect\nu ^{\ast
}=0.167$, $M=2$; (d) $\protect\nu ^{\ast }=0.206$, $M=2$; (e) $\protect\nu %
^{\ast }=0.264$, $M=3$; (f) $\protect\nu ^{\ast }=0.380$, $M=4$. }
\label{fig4}
\end{figure}

We remark that our calculation does not include disorder so that the
lowest-energy mode in figs. 4 is a gapless phonon mode. For small wave
vector, this phonon mode has the typical $\omega \sim k^{3/2}$ dispersion of
a Wigner crystal. In the presence of disorder, this phonon mode would be
gapped.

Upon entering a state with $M$ electrons per bubble, the dispersion and
maximal frequency of the phonon mode first increases with the filling
fraction. Near the critical filling factors $\nu _{M\rightarrow M+1}^{\ast }$
computed above in the HFA, the phonon mode starts to soften at a finite wave
vector $\mathbf{k}$. This indicates that the bubble states become locally
unstable when the outer rings of adjacent bubbles start to touch. For the
cases $N=2,3,5$ that we have studied, we find that the phonon mode frequency
vanishes at this wavevector for $\nu ^{\ast }>$$\nu _{M\rightarrow
M+1}^{\ast }$ so that a second order transition is preempted by a first
order one. This is also the case for the transition between the $M=N+1$
bubble state and the stripe state as we show in Fig. 5 where we plot the
dispersion relation of phonon mode of the stripe state for wave vector $k$
along the direction of the stripes ($k_{x}=\pi/\xi$ where $\xi$ is the
interstripe distance). The dispersion is shown for several values of the
filling factor from $\nu ^{\ast }=0.42$ to $\nu ^{\ast }=0.50$ in Landau
level $N=2$. It is clear that the dispersion of the phonon (and in
particular the region around the roton miminum) does not change
significatively near $\nu =0.428,$ the filling factor at which the
Hartree-Fock calculation predicts a transition between the $M=3$ bubble
state and the stripe state.

\begin{figure}[tbp]
\includegraphics[width=8.25cm]{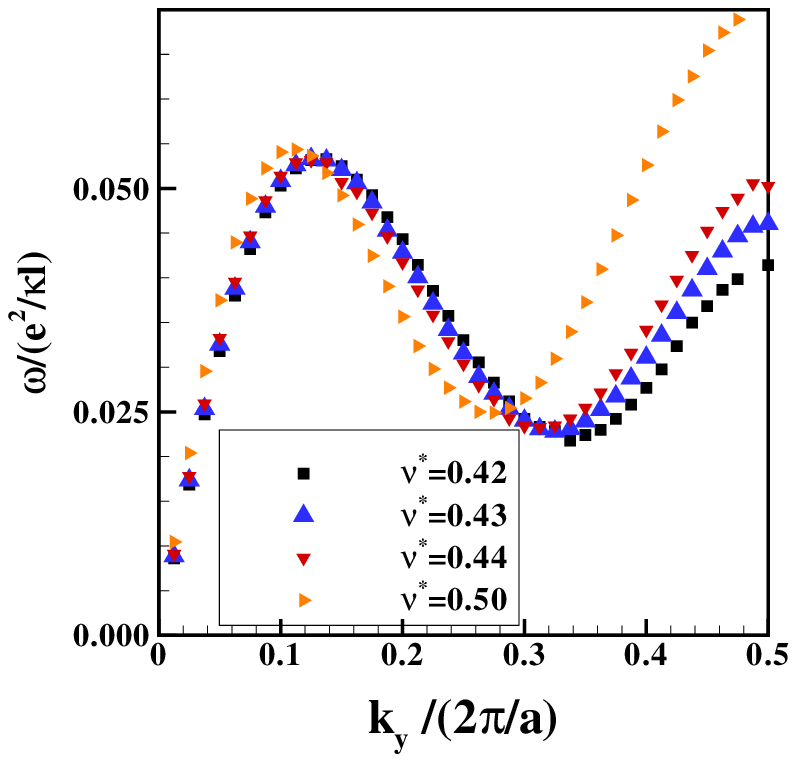}
\caption{Dispersion relation of the phonon mode of the modulated stripe
state in Landau level $N=2$ for wave vector $\mathbf{k}$ along the direction
of the stripes ($k_{x}=\protect\pi/\protect\xi$). Here, $a$ is the period of
the modulations along the stripes and $\protect\xi$ is the interstripe
distance. }
\label{fig5}
\end{figure}

For Landau level index $N=1$, we find an interesting exception in that the
softening of the phonon mode occurs for $\nu ^{\ast }<\nu _{1\rightarrow
2}^{\ast }$. This indicates the possible existence of some new charge
ordered state. We did not study this case further, however, since our
approximations do not include correlations responsible for the correlated
liquid states that are present in $N=1$\cite{yoshioka}.

Another indication of the limitations of the HFA and TDHFA is that the last
bubble state ($M=N+1$) is stable, in contrast with the DMRG result\cite%
{yoshioka}. The DMRG method includes correlations neglected in the HFA and
is, in principle, more exact than the HFA. In any case, for values of $\nu $
where a bubble state is stable, we expect the TDHFA to give a reasonable
approximation of the dispersion of its collective excitations.

We see from figs. 4 that, above the phonon mode and in the region where the $%
M=1$bubble state is stable, there is a series of higher-energy modes that
are almost dispersionless except near filling factors where the transitions
between bubble states do occur. As we explained in Ref. \onlinecite{cote1},
more and more of these modes appear in the response function when we
increase the size of the matrix $M_{\mathbf{G},\mathbf{G}}\left( \mathbf{k}%
\right) $in the equation of motion (Eq. (\ref{7p1})) for the response
function $\chi _{\mathbf{G},\mathbf{G}^{\prime }}\left( \mathbf{k},\omega
\right) $. As more modes appear, the previous ones are not shifted in energy
so that these modes are not numerical artifacts associated with the
truncation of the infinite dimensional matrix $M_{\mathbf{G},\mathbf{G}%
}\left( \mathbf{k}\right) $. For higher values of $M$, a general trend of
the dispersion (except close to transition points) is that there are $M$
low-energy dispersive modes close in energy and a number of higher-energy
much less dispersive modes. For large values of $M$, each bubble has a
complex pattern of density modulations and a large number of reciprocal
lattice vectors are needed to describe its structure adequately. It follows
that our numerical procedure is less precise in this case as can be seen
from Fig. 4(f).

The small dispersion of the higher-energy modes suggests that they can be
identified as local oscillations of the density within a bubble. This is
indeed what can be seen from an animation of these modes using Eq. (\ref%
{1_13}). Figs. 6 show several snapshots of these modes representing the
motion of the density in the second, third, and fourth modes for $M=1$ in
Landau level $N=2$. From these figures, we see that the higher-energy modes
are density waves propagating along the rings of the bubbles. These modes
can be readily identified by the number of wavelengths of these waves
enclosed in the perimeter of the rings. For example, the second mode
corresponds to the case where the perimeter of the rings enclosed two
wavelengths, the third mode to the case where the perimeter enclosed three
wavelengths and so on. When the separation between the bubbles decreases in
a given $M$ bubble state, the density waves on each bubble become more and
more coupled and the dispersion of the higher-energy modes can be very
pronounced.

\begin{figure}[tbp]
\includegraphics[width=8.25cm]{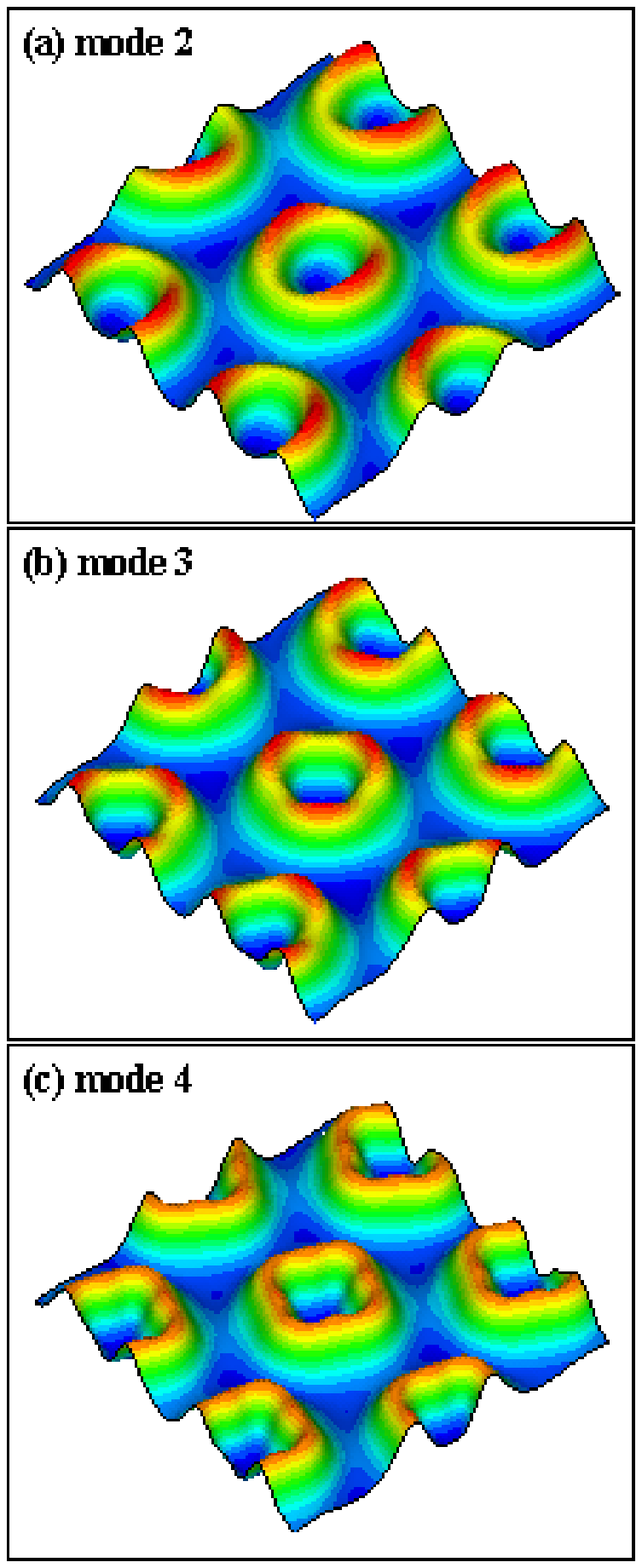}
\caption{Snapshots of the motion of the density in the second, third, and
fourth higher-energy collective modes of the $M=1$ bubble state in Landau
level $N=2$. The filling factor is $\protect\nu ^{\ast }=0.113$. (See
fig6.png file)}
\label{fig6}
\end{figure}

The frequency of the higher-energy collective modes change discontinuously
with the transition between the bubble states. A measurement of the
frequency of these modes could then provide yet another signature of the
bubble states. It would be interesting if variations in the gap $\omega
\left( \mathbf{k}\rightarrow 0\right) $of these higher-energy modes could be
detected experimentally. Recently, Lewis et al.\cite{lewis} reported the
observation of an absorption peak in a measurement of the microwave
absorption of a ultrahigh mobility two-dimensional electron system in Landau
levels $N=2$ and $N=3$. A peak of absorption occurs at a frequency that
decreases as the center ($\nu ^{\ast }=0.5$) of the Landau level is
approached from either above or below. The resonances in the absorption are
sharpest for filling factors $\nu ^{\ast }$ around $1/4$ and $3/4$
corresponding to the values where the RIQHE occur in transport experiments%
\cite{lilly}. At these filling factors, the resonance frequency is
approximately $f=500$MHz. It is natural to associate this resonance with a
pinned phonon mode of the bubble states (of either electrons or holes for $%
\nu ^{\ast }<0.5$or $\nu ^{\ast }>0.5$).

\begin{figure}[tbp]
\includegraphics[width=8.25cm]{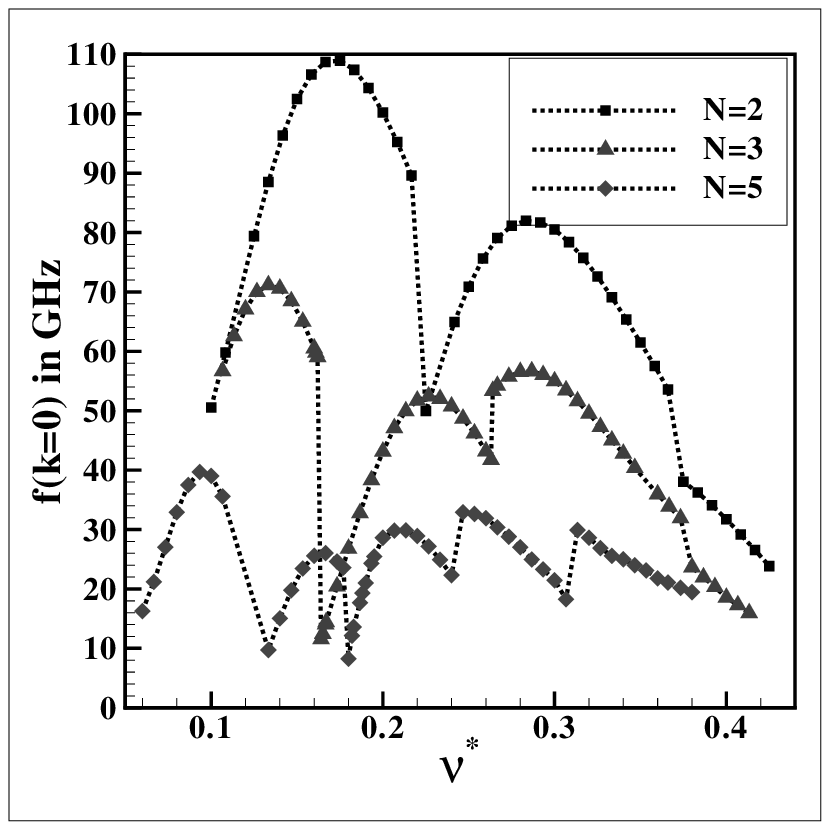}
\caption{Frequency $f\left( \mathbf{k}=0\right) $ of the first higher-energy
collective mode of the bubble states versus filling factor $\protect\nu %
^{\ast }$ in Landau levels $N=2,3,5.$ We have assumed a density of $%
n=3.2\times 10^{11}$ e/cm$^{2}$.}
\label{fig7}
\end{figure}

The higher-energy modes of the bubble states are gapped at $\omega \left( 
\mathbf{k}=0\right) =0$. In Fig. 7, we plot the gap $f\left( \mathbf{k}%
=0\right) =\omega \left( \mathbf{k}=0\right) /2\pi $ in the first
high-energy mode as a function of filling factor $\nu ^{\ast }$ for Landau
levels $N=2,3,5$. To convert our frequencies (in units of $e^{2}/\hslash
\kappa \ell $) to GHz, we assume $\kappa =12.9$ for the dielectric constant
of the host semiconductor (GaAs) and take a typical density of $n=3.2\times
10^{11}$ electrons/cm$^{2}$ so that $e^{2}/\hslash \kappa \ell =3827/\sqrt{%
\nu }$GHz where $\nu $ is the \textit{total} filling factor of the 2DEG. At
filling factor $\nu $, the magnetic field is given by $B=nh/e\nu =13.2/\nu $
Tesla. As Fig. 7 shows, the transitions between bubble states lead to abrupt
changes in the gap frequency when the filling factor is varied. The
discontinuities are more pronounced for lower Landau level indices $N$. The
frequency range of the first higher-energy mode is in a range of frequencies
that can be obtained in a microwave absorption experiment; thus, such
experiments can in principle proble the transitions among bubble states. We
remark that other effects not included in our calculation, particularly the
modeling of a finite width for the electronic wave function in the quantum
well, could lead to a reduction of the gap frequency. This would make
behavior such as that seen in Fig. 7 easier to detect.

To complete our calculation, it would be necessary to find out whether these
higher-energy collective excitations are really detectable by a measurement
of the absorption power. For this, one needs to compute the longitudinal
conductivity $\sigma _{xx}$in the presence of the disorder potential since
otherwise (by Kohn's theorem) only the cyclotron mode will show up in $%
\sigma _{xx}$. This calculation is beyond the scope of this paper but is in
progress and will be reported elsewhere.

\section{Conclusion}

In this work, we have studied several physical properties of the bubble
states that form in higher Landau levels. We have computed the energy and
density pattern of the 2DEG ground state, in Landau levels $N=2,3,$and $5$,
when the filling factor of the partially filled level is gradually increased
from $\nu ^{\ast }=0$ to $\nu ^{\ast }=0.5$. In the Hartree-Fock
approximation, the Wigner crystal at small $\nu ^{\ast }$evolves into the
modulated stripe state near $\nu ^{\ast }=0.5$ by passing through a
succession of bubble states with increasing number of guiding centers per
bubble. In all cases ($N>1$) that we have studied, the transitions are first
order.

We have shown that several physical quantities such as the single-particle
density of states, the magnetization, the magnetic susceptibility and the
dispersion of the collective excitations change discontinuously at the
transition between these different ground states. We believe that these
abrupt changes can be detected experimentally. In particular, we noticed
that the density of states structure in the bubble states has features that
allow one to determine, at least in principle, the number of guiding centers
per bubble.

We have studied more closely the collective excitations of the bubble states
and, in particular, the structure of the higher-energy modes (i.e. those
modes above the phonon mode). In light of the recent microwave experiments
by Lewis et al.\cite{lewis}, we think that some of these modes are likely to
be accessible experimentally. More work is needed to compute the real weight
of these modes in the absorption. If these higher-energy modes can be
detected in microwave absorption experiments, they will show a discontinuous
change of the frequency $f\left( \mathbf{k}=0\right) $ at each transition
between the bubble states.

\section{Acknowledgements}

The authors thank R. Lewis and L. Engel for several useful discussions. This
work was supported by a research grant (for R.C.) and undergraduate research
grants (for C.D. and J.B.) from the Natural Sciences and Engineering
Research Council of Canada (NSERC). H.A.F. acknowledges the support of the
NSF through Grant No. DMR-0108451.

\end{document}